\documentclass[preprint,amsmath,amssymb,prb]{revtex4-1}
\usepackage{graphicx}
\usepackage{dcolumn}
\usepackage{bm,bbm}
\usepackage{color}
\usepackage[colorlinks=true,linkcolor=blue]{hyperref}
\usepackage{ulem}
\usepackage{comment}

\newcommand{\bZ}{\mathbb{Z}}

\newcommand{\cH}{\mathcal{H}}
\newcommand{\cJ}{\mathcal{J}}
\newcommand{\cO}{\mathcal{O}}

\newcommand{\U}{\mathrm{U}}

\newcommand{\SU}{\mathrm{SU}}

\newcommand{\fg}{\mathfrak{g}}

\newcommand{\beq}{\begin{eqnarray}}
\newcommand{\eeq}{\end{eqnarray}}
\newcommand{\bea}{\begin{eqnarray}}
\newcommand{\eea}{\end{eqnarray}}		  

\begin{document}

\title{Fermi/non-Fermi Mixing in SU($N$) Kondo Effect}

\author{Taro Kimura$^{1,2}$}%
 \email{taro.kimura@keio.jp}
\author{Sho Ozaki$^1$}%
 \email{sho.ozaki@keio.jp}
 
\affiliation{%
 ${}^1$Department of Physics \& Research and Education Center for Natural Sciences,
 Keio University, Kanagawa
 223-8521, Japan}
 \affiliation{%
 ${}^2$Fields, Gravity \& Strings, CTPU, Institute for Basic Science, Daejeon 34047, Korea}

 \begin{abstract}
  We apply conformal field theory analysis to the $k$-channel SU($N$) Kondo system, and find a peculiar behavior in the cases $N > k > 1$, which we call Fermi/non-Fermi mixing:
  The low temperature scaling is described as the Fermi liquid, while the zero temperature infrared fixed point exhibits the non-Fermi liquid signature.
  We also show that the Wilson ratio is no longer universal for the cases $N > k > 1$.
  The deviation from the universal value of the Wilson ratio could be used as an experimental signal of the Fermi/non-Fermi mixing.
 \end{abstract}

\maketitle
\tableofcontents


\section{Introduction and Summary}
\label{sec:intro}

The Kondo effect has been providing a lot of theoretical and experimental occasions, which deepen our understanding of many-body correlated electron systems.
After the increasing resistivity in the low temperature regime was clarified by using the standard perturbation analysis~\cite{Kondo:1964nea}, the focus has been moved to deeper understanding of the Kondo system beyond the perturbation theory.
Indeed the formation of the Kondo singlet in the regime below the characteristic scale, which is called the Kondo temperature $T_\text{K}$, shows the non-perturbative nature due to the asymptotic freedom of the Kondo system.
This means that we have to rely on a non-perturbative approach to study the low energy physics of the Kondo effect.
This kind of attempts for the non-perturbative study was initiated by Wilson's numerical renormalization group analysis~\cite{Wilson:1974mb}, and then a lot of works appeared following that, including Bethe ansatz~\cite{Andrei:1980fv,Wiegmann:1980JETP}, large $N$ analysis~\cite{Coleman:1983PRB,Read:1983JPC,Coleman:1987PRB,Parcollet:1997PRL,Parcollet:1998PRB}, and so on.
One of the most important non-perturbative approaches to the Kondo problem is conformal field theory (CFT) analysis initiated by Affleck and Ludwig~\cite{Affleck:1990zd,Affleck:1991tk,Affleck:1990by,Affleck:1990iv,Affleck:1992ng,Ludwig:1994nf}.
See also a review article~\cite{Affleck:1995ge}.

In this paper we study the $k$-channel SU($N$) Kondo effect based on the CFT approach.
In particular, we put a special focus on the cases $N > k > 1$, which Affleck and Ludwig do not investigate in detail.
We first compute the $g$-factor associated with the impurity entropy for the multi-channel system~\cite{Affleck:1991tk}.
Since the $g$-factor becomes a rational (irrational) value for the Fermi (non-Fermi) liquid fixed point, it can be used to clarify the infrared (IR) fixed point of the Kondo system.
We then consider the low temperature scaling of the specific heat and susceptibility, which can be studied by the conformal perturbation theory around the corresponding IR fixed point.
This perturbation analysis is performed with respect to the corresponding leading irrelevant operator the explicit form of which depends on the number of the channel $k$. 

We obtain the the irrational values of the $g$-factor in the cases $N > k > 1$, which imply the non-Fermi liquid fixed point. 
On the other hand, the low temperature analysis shows that the scaling behavior coincides with that of the Fermi liquid in the cases $N > k > 1$.
We call this peculiar behavior Fermi/non-Fermi mixing, the low temperature scaling shows the Fermi liquid behavior, while its IR fixed point shows the non-Fermi liquid signature in $N > k > 1$.
We remark that such a behavior was pointed out before based on the Bethe ansatz~\cite{Jerez:1998PRB}, and a similar CFT analysis~\cite{Parcollet:1997PRL,Parcollet:1998PRB}.
We summarize the dependence on the fundamental parameters $(N,k)$ in Table~\ref{Tab:Fermi/non-Fermi}.
We also find that in the cases $N > k > 1$ the Wilson ratio is no longer an universal quantity of the Kondo system, which depends on the detail of the system. 
In this paper we basically focus on the impurity in the fundamental representation of $\SU(N)$.
Therefore we do not enter the underscreening regime even in the case $N > k > 1$.

\begin{table*}[t]
 \centering
 \begin{tabular}{c@{\hspace{3em}}c@{\hspace{3em}}c}\hline\hline
  & $k \ge N > 1$ & $N > k > 1$ \\\hline  
  IR fixed point (ground state) & non-Fermi & non-Fermi \\ 
  Low $T$ scaling (excitation) & non-Fermi & Fermi \\
  Wilson ratio & non-Fermi (universal) & mixed (non-universal)
  \\\hline\hline
 \end{tabular}
 \caption{Fermi and non-Fermi liquid behavior of $k$-channel SU($N$) Kondo system at the IR fixed point and in the low temperature regime. We observe the Fermi/non-Fermi mixing in the regime $N > k > 1$, whose low temperature scaling is described as the Fermi liquid, while its IR fixed point is of the non-Fermi liquid. Correspondingly the Wilson ratio becomes non-universal in the latter case.}
 \label{Tab:Fermi/non-Fermi}
\end{table*}

The remaining part of this paper is organized as follows.
In Sec.~\ref{sec:CFT} we summarize some basic aspects of the CFT approach to Kondo effect, in particular, characterized by the fundamental parameters ($N, k$) in the system.
In Sec.~\ref{sec:g} we consider the $g$-factor associated with the impurity entropy, which characterizes the IR fixed point of the Kondo system.
We will see that the $g$-factor shows the non-Fermi liquid behavior in general for the multi-channel systems $(k>1)$.
We also provide an interpretation of the $g$-factor as the Wilson loop contribution.
In Sec.~\ref{sec:lowT} we compute the low temperature scaling behaviors of the specific heat and susceptibility.
In particular we observe that the scaling behavior is described by the Fermi liquid in the cases $N > k > 1$ although the $g$-factor shows the non-Fermi liquid behavior, which implies the Fermi/non-Fermi mixing.
We also show that, as a result of this mixing, the Wilson ratio doesn't play a role as the universal quantity in $N > k > 1$.
In Sec.~\ref{sec:discussion} we conclude this paper with some remarks and discussion.

\section{Conformal Field Theory Approach to Multi-channel SU($N$) Kondo Effect}
\label{sec:CFT}

Let us briefly summarize the CFT approach to the multi-channel SU($N$) Kondo effect~\cite{Affleck:1995ge}.
The Kondo problem is defined as a (3+1)-dimensional model of the conduction electron interacting with the localized magnetic impurity.
Assuming the impurity is sufficiently dilute, we apply the $s$-wave approximation to the (3+1)-dimensional model.
This approximation leads to the effective (1+1)-dimensional $k$-channel SU($N$) Kondo model as
\beq
H
&=& \frac{1}{2\pi} \left[ \int^{\infty}_{0} dx \, \mathcal{H}_{0} (x) + H_{\rm{K}} \right],
\label{Kondo_H}
\eeq
where the free fermion Hamiltonian density is given by
\beq
\mathcal{H}_{0}
&=& i \psi_{L}^{ i \alpha \dagger} (x) \frac{ \partial \psi_{L i \alpha} (x) }{ \partial x }
- i \psi_{R}^{i \alpha \dagger} (x) \frac{ \partial \psi_{R i \alpha} (x) }{ \partial x },
\label{free_H}
\eeq
and the Kondo interaction term is
\beq
H_{\rm{K}}
&=& \frac {\pi}{ 2} \lambda_{\rm{K}} S^{a} \left( \psi_{L}^{i \alpha \dagger} (0) + \psi_{R}^{i \alpha \dagger} (0) \right) \, \left( t^{a} \right)_{\alpha \beta} \, \left( \psi_{L i}^{\beta} (0) + \psi_{R i }^{\beta} (0)  \right).
\eeq
Here, $\psi_{R}$ and $\psi_{L}$ are right and left moving fermions, which have SU($N$) spin and SU($k$) channel (flavor) indices $\alpha, \beta = 1, \cdots, N$ and $i = 1, \cdots, k$, respectively.
$t^{a}$ ($a= 1, \cdots, N^{2} - 1$) are the generators of SU($N$) group, and $S^{a}$ is a local SU($N$) spin at the origin $x=0$.
The fermion fields satisfy the boundary conditions $\psi_{L}(0) = \psi_{R} (0)$ and $\psi_{L}(x) = \psi_{R}(-x)$, owing to the $s$-wave approximation~\cite{Affleck:1990iv}.
Therefore, we can express the Hamiltonian (\ref{Kondo_H}) in terms of only the left moving fermion fields as
\beq
H
&=& \int^{\infty}_{-\infty} dx \, \left\{ \frac{1}{2\pi} i \psi_{L}^{i \alpha \dagger} (x) \frac{ \partial \psi_{L i \alpha} (x) }{ \partial x } + \lambda_{\rm{K}} \psi_{L}^{i \alpha \dagger} (x) (t^{a})_{\alpha \beta} \psi_{L i}^{\beta} (x) S^{a} \delta(x) \right\}.
\eeq
Hereafter, we shall drop the subscript $L$. The kinetic term (first term) is equivalent to the
Wess--Zumino--Witten (WZW) model whose action, associated with the Kac--Moody algebra $\widehat{\fg}_k$, is given by
\begin{align}
 S_k[g] & =
 \frac{k}{16\pi} \int d^2 x \,
 \partial_\mu g \partial^\mu g^{-1}
 + \frac{k}{24\pi} \int d^3 x \,
 \epsilon^{\mu\nu\lambda}
 \left( g^{-1} \partial_\mu g \right)
 \left( g^{-1} \partial_\nu g \right)
 \left( g^{-1} \partial_\lambda g \right)
\end{align}
where $g$ is an element of group $G$ associated with the Lie algebra $\fg$, and $k$ is the level corresponding to the number of the channel.
In particular, for the $\SU(N)$-spin $k$-channel Kondo model, the total action of the kinetic term consists of three parts
\begin{align}
 S_k[g \in \SU(N)] + S_N[h \in \SU(k)] +
 \frac{Nk}{2} \int d^2 x \, (\partial_\mu \phi)^2
\end{align}
with the symmetry
\begin{align}
 \widehat{\SU(N)}_k \times \widehat{\SU(k)}_N \times
 \widehat{\U(1)}_{Nk} 
 \label{eq:sym_Kondo}
 \, .
\end{align}
The bosonic field $\phi(x)$ describes the $\U(1)$ degrees of freedom.
They correspond to the spin, channel, and charge part, respectively.
This factorization is nothing but the spin-charge separation.
Indeed, defining the spin, channel, and charge currents
\beq
J^{a} (x)
&=& : \psi^{i \alpha \dagger} (x) ( t^{a})_{\alpha \beta} \psi_{i}^{\beta} (x) :, \\
J^{A} (x)
&=& : \psi^{i \alpha \dagger} (x) ( T^{A})_{i j} \psi_{\alpha}^{j} (x) :, \label{FlavorCurrent} \\
J (x)
&=& : \psi^{i \alpha \dagger} (x)  \psi_{i \alpha } (x) :,
\eeq
we can express the Hamiltonian density in the Sugawara form as
\beq
\mathcal{H}
&=& \frac{1}{(N+k)} : J^{a} J^{a} (x): + \frac{1}{(k+N)} : J^{A} J^{A}(x): + \frac{1}{2 N k } :J^{2}(x):
+ \lambda_{\rm{K}} J^{a} S^{a} \delta(x), \nonumber \\
 \label{eq:Kondo_Ham1} 
\eeq
where $: OO(x): \ = \lim_{\epsilon \to 0} \left\{ O(x) O(x+ \epsilon) - \langle O(x) O(x+ \epsilon) \rangle \right\}$ is the normal order product, which subtracts the singularities at $\epsilon \to 0$.
$T^{A}$ $(A = 1, \cdots, k^{2}-1)$ in Eq.~(\ref{FlavorCurrent}) are the generators of the flavor (channel) SU($k$) symmetry group.
We see that the Hamiltonian density (\ref{eq:Kondo_Ham1}) consists of spin, channel and charge parts with the Kondo interaction. 
The SU($N$) spin current $J^{a}(x)$ satisfies the Kac--Moody algebra $\widehat{\fg}_k$:
\beq
\left[ J^{a}(x), J^{b}(y) \right]
&=& i f^{ab}_{~~c} J^{c}(x) \delta(x-y) + \frac{i}{2\pi} \left( \frac{k}{2} \right) \delta^{ab} \frac{ \partial }{\partial x } \delta(x-y),
\eeq
where $f^{ab}_{~~c}$ is the structure constant, while the Fourier mode of the current
\beq
J^{a}_{n}
&=& \int^{L}_{-L} dx \, {\rm{e}}^{ i n \frac{ \pi }{ L } x } J^{a}(x)
\eeq
obeys 
\begin{align}
 \left[ J_n^a, J_{m}^b \right]
 & =
 i f^{ab}_{~~c} J_{n+m}^c + \frac{k}{2} n \delta^{ab} \delta_{n+m,0} \, .
\end{align}
Here $L$ in the Fourier integral stands for the size of the space, which is taken to be $L \to \infty$.
Now, the impurity is a singlet with respect to the channel and charge symmetry $\SU(k) \times \U(1)$, while the electron is in the fundamental representation of spin and channel $\SU(N) \times \SU(k)$.
We need to specify the representation of the impurity $R_\text{imp}$ under the spin $\SU(N)$ symmetry, which characterizes the Kondo model as the fundamental parameter $(N,k,R_\text{imp})$.

We can complete the square for the Hamiltonian \eqref{eq:Kondo_Ham1}, up to a constant term $S^a S^a$,
\begin{align}
 \cH & =
 \frac{1}{(N+k)} \cJ^a \cJ^a
 + \frac{1}{(k+N)} J^A J^A
 + \frac{1}{2 Nk} J^2
 \label{eq:Kondo_Ham2} 
\end{align}
with the current
\begin{align}
 \cJ^a = J^a + \pi (N+k) \, \lambda_\text{K} \, \delta(x) \, S^a
\end{align}
which obeys the same current algebra $\widehat{\mathfrak{su}(N)}_k$ if the coupling constant takes a special value $\lambda_\text{K}^* := 2/(N+k)$. The coupling value $\lambda_\text{K}^*$ is regarded as a renormalized coupling at the IR fixed point.
This redefinition can be interpreted as absorbtion of the impurity spin at the IR fixed point, and the corresponding Hamiltonian \eqref{eq:Kondo_Ham2} has the same symmetry \eqref{eq:sym_Kondo} as the original model \eqref{eq:Kondo_Ham1} in the ultraviolet (UV) region.
The impurity effect at the IR fixed point is then implemented as the boundary condition of the corresponding CFT.
From the $\SU(N)_k$ fusion rule, we can obtain the spectrum of the system by studying how the highest weight states fuse with the boundary operator describing the impurity spin.



\section{Impurity Entropy: $g$-factor}
\label{sec:g}

We can exactly compute the impurity contribution to thermodynamic entropy at the IR fixed point using a ratio of the modular $S$-matrix,\footnote{
The modular $S$-matrix formula for $\SU(N)$ theory is given by
\begin{align*}
 S_{\mu\nu} & =
 q^{-\frac{1}{N} c(\mu) c(\nu)}
 \det_{1 \le i, j \le N} q^{(\mu_i + N - i)(\nu_j + N - j)}
\end{align*}
where $\mu$ and $\nu$ are partitions characterizing the $\SU(N)$ representation, e.g. $\mu = (\mu_1, \mu_2, \ldots, \mu_N) \in \bZ_{\ge 0}^N$, obeying $\mu_1 \ge \mu_2 \ge \cdots \ge \mu_N \ge 0$.
The parameter $q$ is defined as $q = \exp \left( 2 \pi i /(N+k)\right)$ for $\SU(N)_k$ theory.
See, for example, Ref.~\onlinecite{DiFrancesco:1997nk}.
In this paper, we use the same notation for the partition and representation as long as no confusion.
The constant $c(\mu)$ is defined as $\displaystyle c(\mu) = \sum_{i=1}^N (\mu_i + N - i)$.
}
which is a logarithm of the so-called $g$-factor~\cite{Affleck:1991tk}
\begin{align}
 S_\text{imp} & = \log g(R_\text{imp})
 \qquad \text{with} \qquad
 g(R) = \frac{S_{0R}}{S_{00}}
 \, .
 \label{eq:g-fac}
\end{align}
The modular $S$-matrix is related to the boundary condition of the corresponding CFT, implying the impurity effect at the IR fixed point. Here $R_{\rm{imp}}$ is the representation of the impurity.
To understand the physical meaning of the $g$-factor, let us consider the standard single channel SU($2$) spin Kondo effect as a simple example.
When the IR fixed point is well described by the local Fermi liquid, the factor can be given as $g = 2 s + 1$ for the impurity spin $s$.
In UV regions, the impurity spin is $s = 1/2$ and thus $g = 2$, while
at the IR fixed point the Kondo singlet state leads to $s \to 0$ and then $g \to 1$, which is consistent with the $g$-theorem claiming that the $g$-factor monotonically decreases in the renormalization flow~\cite{Affleck:1991tk}.
This is a standard behavior of the $g$-factor at the IR fixed point in Fermi liquid with the spin $\SU(2)$ symmetry.
However, in general, this $g$-factor takes an irrational value, which suggests the non-Fermi liquid behavior at the IR fixed point, typically observed in the over-screening Kondo system, while we obtain an integer $g$ for critical and under-screening Kondo systems which are described as the local Fermi liquid.%

\subsection{Relation to Wilson Loop}

We remark the $g$-factor formula \eqref{eq:g-fac} is exactly the same as the (unknot) Wilson loop average $\left\langle W (R) \right\rangle$ of $\SU(N)_k$ Chern--Simons theory in the representation $R$~\cite{Witten:1988hf}, %
\begin{align}
 \left\langle W (R) \right\rangle = \frac{S_{0R}}{S_{00}}
 \, .
 \label{eq:Schur}
\end{align}
The right hand side of Eq. (\ref{eq:Schur}) can be expressed in terms of the quantum dimension of $R$,
\begin{align}
\frac{S_{0R}}{S_{00}} =  \operatorname{dim}_q R
 \, .
\end{align}
This leads to
\begin{align}
 g(R) = \left\langle W (R) \right\rangle = \operatorname{dim}_q R
  \, .
\end{align}
We have a formula to compute this quantum dimension~\cite{DiFrancesco:1997nk}
\begin{align}
 \operatorname{dim}_q R
 & =
 \prod_{i<j}^N
 \frac{
 q^{\frac{1}{2}(R_i - R_j - i + j)} - q^{-\frac{1}{2}(R_i - R_j - i + j)}}
 {q^{\frac{1}{2}(- i + j)} - q^{-\frac{1}{2}(- i + j)}}
 \, .
 \label{eq:q-dim}
\end{align}
where $q = \exp\left( 2\pi i /(N+k) \right)$ for $\SU(N)_k$ theory and
we use the notation that $R = (R_1, \ldots, R_N) \in \bZ^N_{\ge 0}$ also denotes the partition characterizing the $\SU(N)$ representation, obeying $R_1 \ge \cdots \ge R_N \ge 0$.
See Appendix~\ref{sec:Schur} for the derivation of the formula \eqref{eq:q-dim}.
The quantum dimension \eqref{eq:q-dim} is reduced to the ordinary dimension of the representation $R$ in the limit $q \to 1$, corresponding to the large channel limit $k \to \infty$.
The Wilson loop appearing here is interpreted as the world line of the impurity located at the boundary, and the $g$-factor corresponds its expectation value.
This implies that the spin absorbtion at the IR fixed point is realized as 
an appearance 
of the Wilson loop.

\subsection{Analysis of $g$-factor}

We compute the $g$-factor for the $k$-channel SU($N$) Kondo system using the formula \eqref{eq:q-dim}.
First of all, for the single-channel system $k = 1$, we shall show that the $g$-factor becomes trivial for arbitrary $N$.
Since the representation of SU($N$)$_k$ has to satisfy $R_i \le k$ for $i = 1, \ldots, N$, a possible representation for $k = 1$ is given by
\begin{align}
 R =
 ( \underbrace{1, \ldots, 1}_{r}, \underbrace{0, \ldots, 0}_{N-r})
\end{align}
which is the $r$-th antisymmetric representation.
In this case, we can directly compute the formula \eqref{eq:q-dim} with arbitrary $r \in \{0, \ldots, N \}$ as
\begin{align}
 \operatorname{dim}_q R = 1
 \quad \text{for} \quad
 k = 1
 \, .
\end{align}
This trivial value implies that the IR fixed point of the single-channel ($k=1$) Kondo system is described as the local Fermi liquid for arbitrary $N$.
The localized spin is completely screened, and forms the SU($N$) Kondo singlet with the conduction electron in this case.

Next we compute the quantum dimension for $k > 1$ and the fundamental representation of the impurity, corresponding to $R=(1,0,\ldots,0)$, which is given by
\begin{align}
 \operatorname{dim}_q \mathbf{N}
 & = \left[ N \right]_q
 = \sum_{i=1}^N q^{\rho_i}
 \label{eq:q-dim_fund}
\end{align}
where $\rho$ is the Weyl vector \eqref{eq:Weyl_vec}, and the $q$-number $[x]_q$ is defined by
\begin{align}
 [x]_q & =
 \frac{q^{x/2} - q^{-x/2}}{q^{1/2} - q^{-1/2}}
 \, .
\end{align}
This $q$-number is reduced to the ordinary number $[x]_q \to x$ in the limit $q \to 1$, so that $\operatorname{dim}_q R \to \operatorname{dim} R$.
Therefore, in this limit, the quantum dimension becomes
\beq
\operatorname{dim}_q \mathbf{N}
 \ \stackrel{q \to 1}{\longrightarrow} \
 N.
\eeq
We remark the anti-fundamental representation $\bar{\mathbf{N}}$ gives the same quantum dimension.

Let us consider the large $N$ behavior of the $g$-factor.
Expanding the expression \eqref{eq:q-dim_fund} with respect to the large $N$ at a fixed $k$, we obtain
\begin{align}
 g & =
 k - \frac{k(k^2-1)}{N^2} \frac{\pi^2}{6} + O(N^{-3})
 \, .
\end{align}
In the large $N$ limit, the $g$-factor is approximated to $g = k$, and the correction starts with $O(N^{-2})$.
This implies that the SU($N$)$_k$ Kondo effect is described as the Fermi liquid in the large $N$ limit, and thus the low-temperature scaling of the specific heat and so on is expected to exhibit the Fermi liquid behavior.

\begin{table}[t]
 \centering
 \begin{tabular}{c@{\qquad}cccc@{\qquad}c}\hline\hline
  $N$ & $k=1$ & $k=2$ & $k=3$ & $k=4$ & $k=\infty$\\ \hline
  2 & 1 & 1.4142... & 1.6180... & 1.7320... & 2 \\
  3 & 1 & 1.6180... & 2 & 2.2469... & 3 \\
  4 & 1 & 1.7320... & 2.2469... & 2.6131... & 4\\[.3em]
  $\infty$ & 1 & 2 & 3 & 4 & $\infty$ \\[.3em]
  \hline\hline
 \end{tabular}
 \caption{Numerical values of the $g$-factor for the (anti)fundamental representation. It approaches to an integer in the large $(N,k)$ limit.}
 \label{table:g-factor}
\end{table}

Table~\ref{table:g-factor} shows the numerical values of the $g$-factor for the (anti)fundamental representation.
Although there is an accidental case giving an integer value for three-channel SU(3) system SU(3)$_{k=3}$, we obtain irrational values for $k > 1$ in general cases.
This is a signature of the non-Fermi liquid behavior at the IR fixed point of the multi-channel Kondo system, corresponding to zero temperature (ground state).
We remark that the coincidence of $g$-factor for SU($N$)$_k$ and SU($k$)$_N$ reflects the level-rank duality of the Kac--Moody algebra.
The SU(3)$_3$ system is self-dual in this sense $(N,k)=(3,3)$.

\section{Low Temperature Behavior}
\label{sec:lowT}

By using the conformal field theory, one can also compute the temperature dependence of several quantities, e.g. specific heat, susceptibility, and resistivity
in the low temperature region compared to the Kondo temperature $T \ll T_\text{K}$.
There are two contributions to these observables, i.e., bulk and impurity contributions.
Let us first consider the bulk contribution to the specific heat, which is obtained by the finite-size scaling argument~\cite{Bloete:1986qm,Affleck:1986bv}
\begin{align}
 C_\text{bulk} & = \frac{\pi}{3} c \, T
 \, ,
 \label{eq:C_bulk}
\end{align}
where the total central charge of the model with the symmetry \eqref{eq:sym_Kondo} is given by
\begin{align}
 c = \frac{k(N^2 - 1)}{N + k} + \frac{N(k^2 - 1)}{k + N} + 1
 & = Nk
 \, .
\end{align}
We remark this bulk contribution does not depend on the impurity representation $R_\text{imp}$.

The bulk susceptibility obtained from the two-point function of the currents is determined by the level of the Kac--Moody algebra~\cite{Affleck:1986sc}
\begin{align}
 \chi_\text{bulk} & = \frac{k}{2\pi}
 \label{eq:chi_bulk}
\end{align}
due to the diagonal part of the current operator product expansion
\begin{align}
 \mathcal{J}^a(z) \mathcal{J}^a(w) & = \frac{k/2}{(z-w)^2} + \text{regular}
 \, .
\end{align}
Here we do not take the summation over the index $a$ on the LHS.
We remark that the first order pole does not appear in this diagonal part, which is proportional to the structure constant $f^{ab}_{~~c} \, \mathcal{J}^c(w) = 0$ for $a = b$.

\subsection{Single-channel System ($k=1$)}

In addition to the bulk contributions, we can compute the low temperature behavior of the impurity part from the conformal field theory.
Firstly, we consider the case of $k=1$. 
In this case, the low temperature scalings of the specific heat and susceptibility are evaluated by Affleck from the conformal field theory \cite{Affleck:1990zd}. For completeness, we here summarize the results of the specific heat and susceptibility for $k=1$.

In the single channel case $k=1$, the scaling behavior at low $T$ is characterized by the leading irrelevant operator $\cO =  \mathcal{J}^a \mathcal{J}^a (x)$ which is the dimension two operator to be localized at the origin.
The perturbation with respect to the leading irrelevant operator
\begin{align}
\delta \mathcal{H}_1 = \lambda_{1} \mathcal{J}^a \mathcal{J}^a (x) \delta (x) 
 \label{eq:pert1}
\end{align}
leads to the impurity contributions to the specific heat and the susceptibility.
Note that from the dimensional analysis, the coupling $\lambda_{1}$ should be proportional to $T_{\rm{K}}^{-1} (\equiv \beta_\text{K})$.
Therefore, in sufficiently low temperatures as compared to $T_{\rm{K}}$, the operator $\delta \mathcal{H}$ becomes irrelevant.
It is difficult to determine the exact form of the coupling $\lambda_{1}$ since it depends on the microscopic details of the system.
However, one can still determine low temperature scaling behaviors of the specific heat and the susceptibility, and the exact form of the Wilson ratio from the fundamental parameters $(N,k)$.
The impurity contributions to specific heat and the susceptibility can be obtained from the first order perturbation of the leading irrelevant operator.
The results are given by \cite{Affleck:1990zd}
\bea
C_{\rm{imp}}
&=& - \lambda_{1} \frac{ k (N^{2}-1 ) }{ 3 } \pi^{2} T,
\label{eq:C_FL}
\eeq
for the specific heat, and
\bea
\chi_{\rm{imp}}
&=& - \lambda_{1} \frac{ k (N+k) }{2},
\label{eq:chi_FL}
\eea
for the susceptibility, with $k=1$.
At the low temperature, the specific heat is proportional to $T$, whereas the susceptibility is independent of the temperature.
These behaviors indicate that the system can be described by the Fermi liquid in the IR regions.
This is consistent with the $g$-factor analysis with $k = 1$ in the previous section.
Gathering the above expressions for the specific heat and the susceptibility, we obtain the Wilson ratio, which is a universal quantity of the Kondo effect, characterizing the IR fixed point, 
\begin{align}
 R_\text{W} & =
 \left( \frac{\chi_\text{imp}}{C_\text{imp}} \right)
 \Bigg/
\left( \frac{\chi_\text{bulk}}{C_\text{bulk}} \right)
 =
 \frac{N}{N-1}.
 \ \ \ \ (k =1)
 \label{eq:Wratio1}
\end{align}
Note that the unknown parameter $\lambda_{1}$ is canceled and does not appear in the Wilson ratio, which is a reason why this ratio is universal.

\subsection{Multi-channel System ($k > 1$)}

Next, we consider cases with $k > 1$, corresponding to the multi-channel system.
In these cases, Affleck--Ludwig \cite{Affleck:1990iv} investigate the specific heat and the susceptibility by using the CFT approach.
However, they especially focus on $k \ge N$ and do not evaluate the observables for $N > k > 1$ in detail. In this subsection, we comprehensively calculate the specific heat and the susceptibility for arbitrary $N$ with $k > 1$, including the cases $N > k > 1$.

For $k > 1$, we can introduce another irrelevant operator, which is given by $\cO = \mathcal{J}_{-1}^a \phi^a (x)$. This operator has the conformal weight $1 + \Delta$, which is a descendent of the adjoint primary $\phi$ with the weight 
\begin{align}
 \Delta & = \frac{N}{N+k}
 \, .
 \label{eq:conf_dim}
\end{align}
For $k > 1$, such an adjoint operator always appears in arbitrary $N>1$.%
\footnote{%
For $k=1$, one cannot have the operator $\cO = \mathcal{J}_{-1}^a \phi^a$ since the adjoint primary, corresponding to $R=(2,1,\ldots,1,0)$, is not generated by the fusion process.
}
Now, we consider the perturbation analysis with respect to the irrelevant operator
\begin{align}
 \delta \cH = \lambda \, \mathcal{J}_{-1}^a \phi^a (x) \delta(x)
 \, .
 \label{eq:pert2}
\end{align}
From the dimensional analysis, the coupling $\lambda$ should be proportional to $T_{\rm{K}}^{ - \Delta }(=\beta_\text{K}^{\Delta})$. Since $\Delta < 1$, this is the leading irrelevant operator for $k > 1$, which is more relevant than the previous irrelevant operator (\ref{eq:pert1}) in the low temperature region.

The observables can be obtained from the partition function with the leading irrelevant operator (\ref{eq:pert2}) as well as an external source term, which is given by
\beq
Z
&=& {\rm{e}}^{- \beta F(T, \lambda, h) } \nonumber \\
&=& \int \mathcal{D} \psi \mathcal{D} \psi^{\dagger}
\ {\rm{e}}^{ - \int^{\beta/2}_{-\beta / 2 } d \tau \int^{L}_{-L} dx \, \mathcal{H} }
{\rm{exp}} \left\{ \int^{\beta/2}_{- \beta/2} d \tau \left[ \lambda \mathcal{J}_{-1}^{a} \phi^{a}( \tau, 0 ) + \frac{h}{2\pi} \int^{L}_{- L} dx \ \mathcal{J}^{3}(\tau, x) \right] \right\} \nonumber \\
&=& Z_{0} \left\langle 
{\rm{exp}} \left\{ \int^{\beta/2}_{- \beta/2} d \tau \left[ \lambda \mathcal{J}_{-1}^{a} \phi^{a}( \tau, 0 ) + \frac{h}{2\pi} \int^{L}_{- L} dx \ \mathcal{J}^{3}(\tau, x) \right] \right\}
\right\rangle 
\eeq
where $F$ is the free energy, and 
the Hamiltonian density $\mathcal{H}$ is given by (\ref{eq:Kondo_Ham2}).
$L$ is a size of the space, which is taken to be $L \to \infty$, and $h$ is the third component of an external 
SU($N$) magnetic field.
Although the generic coupling between the magnetic field and the current is given by $h^a \mathcal{J}^a$, the SU($N$) symmetry allows us to consider a specific term $h^3 \mathcal{J}^3 = h \, \mathcal{J}^3$ without loss of generality.
$Z_0$ is the partition function with $\lambda = h = 0$.
The free energy $F$ can be divided into the bulk and impurity parts as
\beq
F
&=& L f_{\rm{bulk}} + f_{\rm{imp}}.
\eeq
Since the bulk part of the free energy is extensive in $L$, it is proportional to $L$.
On the other hand, the impurity part of the free energy is not extensive in $L$.
The $f_{\rm{bulk}}$ with vanishing $h$ is nothing but the free energy of non-interacting $Nk$ fermions in (1+1)-dimensions, which is well known as
\beq
f_{\rm{bulk}}
&=& - \frac{c}{6} \pi T^{2}.
\eeq
Here the central charge is $c = Nk$, corresponding to the $Nk$ free fermions.
This free energy leads to the bulk specific heat \eqref{eq:C_bulk} from $C_\text{bulk} = - T^2 \partial^2 f_\text{bulk}/\partial T^2$.

\subsubsection{Specific Heat}
\label{sec:C}

We evaluate the temperature dependence of the specific heat at low temperature from the second order perturbation with respect to the leading irrelevant operator for $k>1$.
Expanding the partition function with respect to the coupling $\lambda$ with $h=0$,
we get 
\beq
{\rm{e}}^{ - \beta f_{\rm{imp}} }
&=&  \left\langle {\rm{exp}} \left\{ \int^{\beta/2}_{-\beta/2} d \tau \  \lambda \mathcal{J}_{-1}^{a} \phi^{a} \right\} \right\rangle \nonumber \\
&=& 1 + \frac{\lambda^{2}}{2} \int^{\beta/2}_{-\beta/2} d \tau_{1} \int^{\beta/2}_{-\beta/2} d \tau_{2}
\left\langle \mathcal{J}_{-1}^{a} \phi^{a} ( \tau_{1}, 0 ) \mathcal{J}_{-1}^{b} \phi^{b}( \tau_{2}, 0) \right\rangle + O (\lambda^{3} ),
\eeq
where the bulk part of the free energy is subtracted by $Z_{0}^{-1}$.
Since the one-point function of Virasoro primary field $\mathcal{J}_{-1}^{a} \phi^{a}$ is trivially zero, the expansion starts with $O(\lambda^2)$ term~\cite{Affleck:1990iv}.
Then, we find the impurity part of the free energy 
\beq
- \beta f_{\rm{imp}}
&=& \frac{ \lambda^{2} }{2} \int^{\beta/2}_{-\beta/2} d \tau_{1} \int^{\beta/2}_{-\beta/2} d \tau_{2}
\left\langle \mathcal{J}_{-1}^{a} \phi^{a} ( \tau_{1}, 0 ) \mathcal{J}_{-1}^{b} \phi^{b}( \tau_{2}, 0) \right\rangle ,
\label{free-energy1}
\eeq
up to the second order of $\lambda$. 
In this (1+1)-dimensional system, the correlation functions can be determined by the Kac--Moody algebra and the conformal symmetry.
At finite temperatures, the correlation function is given by \cite{Affleck:1990iv}
\beq
\left\langle \mathcal{J}_{-1}^{a} \phi^{a} ( \tau_{1}, 0 ) \mathcal{J}_{-1}^{b} \phi^{b}( \tau_{2}, 0) \right\rangle
&=& \frac{ ( N + k/2 ) \delta^{ab} \delta^{ab} }{ \left| \frac{ \beta }{ \pi } {\rm{sin}} \frac{ \pi }{ \beta } ( \tau_{1} - \tau_{2} ) \right|^{ 2 (1+ \Delta) } } \nonumber \\
&=& \frac{ (N^{2} - 1) ( N + k/2 ) }{ \left| \frac{ \beta }{ \pi } {\rm{sin}} \frac{ \pi }{ \beta } ( \tau_{1} - \tau_{2} ) \right|^{ 2 ( 1+ \Delta ) } }.
\label{correlation1-T}
\eeq
The power of the correlation function is determined by the conformal weight of the leading irrelevant operators, while the coefficient corresponds to the OPE coefficient. 
This correlation function leads to 
\beq
- \beta f_{\rm{imp}}
&=& \frac{ \lambda^{2} }{2} \int^{\beta/2}_{-\beta/2} d \tau_{1} \int^{\beta/2}_{-\beta/2} d \tau_{2}
\frac{ (N^{2} - 1) ( N + k/2 ) }{ \left| \frac{ \beta }{ \pi } {\rm{sin}} \frac{ \pi }{ \beta } ( \tau_{1} - \tau_{2} ) \right|^{ 2 ( 1+ \Delta ) } } 
\Theta \left( {\rm{sin}} \left( \left| \frac{ \pi }{ \beta } ( \tau_{1} - \tau_{2} ) \right| \right) - \epsilon  \right), \nonumber \\
\eeq
where the integral is regularized by the step function $\Theta$ with the cut-off $\epsilon = a \pi / \beta$.
Here the short distance parameter $a$ is characterized by the Kondo temperature $T_{K}$ as $a \sim \beta_{\rm{K}} = 1/ T_{\rm{K}}$.
Since the integrand is translational invariant and has the symmetry $\tau_{1}  \leftrightarrow \tau_{2}$, the free energy reads
\beq
- \beta f_{\rm{imp}}
&=& \lambda^{2} \beta \int^{\beta/2}_{a} d \tau \frac{ (N^{2} - 1) ( N + k/2 ) }{ \left| \frac{ \beta }{ \pi } {\rm{sin}} \left( \frac{ \pi }{ \beta } \tau \right) \right|^{ 2 ( 1+ \Delta ) } } 
\eeq
Changing the integral variable as $u = {\rm{tan}} ( \pi \tau / \beta )$, we find
\beq
- \beta f_{\rm{imp}}
&=& \lambda^{2} \beta \left( \frac{ \pi }{ \beta } \right)^{ 2 \Delta + 1 }  (N^{2} - 1) ( N + k/2 ) \int^{\infty}_{{\rm{tan}} \epsilon} du \, ( 1 + u^{2} )^{\Delta} u^{ - 2 ( 1 + \Delta) }  \nonumber \\
&=& \lambda^{2} \beta \left( \frac{ \pi }{ \beta } \right)^{ 2 \Delta + 1 }  (N^{2} - 1) ( N + k/2 ) \left\{
\frac{ \epsilon^{ -1 -2 \Delta } }{ 1 + 2 \Delta } + 2 \Delta \int^{\infty}_{\epsilon} du \, ( 1 + u^{2} )^{ \Delta - 1} \frac{ u^{-2 \Delta } }{ 1 + 2 \Delta } \right\}. \nonumber \\
\eeq
The first term is proportional to the inverse temperature $\beta$, and thus does not contribute to the specific heat.
The contribution to the specific heat comes from the second term which includes the the following integral:
\beq
I_{1} (\epsilon)
&=& \int^{\infty}_{\epsilon} du \, ( 1 + u^{2} )^{ \Delta - 1}  u^{-2 \Delta }  \nonumber \\
&=& 
\left\{  \begin{array}{ll}
\displaystyle
 \frac{1}{2} \left( \frac{ 1 - 2\Delta }{2} \right) \frac{ \Gamma(1/2 - \Delta ) \Gamma(1/2) }{ \Gamma(1-\Delta ) }, & \ \ \ \ \ \ (k > N) \\[1.5em]
\displaystyle	  
{\rm{log}} \left( \frac{T_{\rm{K}}}{T} \right) + ({\rm{const.}}), &  \ \ \ \ \ \ (k = N) \\[1.5em]
\displaystyle
\frac{ \epsilon^{-2\Delta + 1} }{ 2 \Delta -1 } + ({\rm{const.}}), & \ \ \ \ \ \ (N > k > 1)
  \end{array} \right.
\eeq
as $\epsilon \to 0$ ($T / T_{\rm{K}} \to 0$).
Then, the free energy is given by
\beq
- f_{\rm{imp}}
&=& \lambda^{2} \left( \frac{ \pi }{ \beta } \right)^{ 2 \Delta + 1 }  (N^{2} - 1) ( N + k/2 ) \frac{ 2 \Delta }{ 1 + 2 \Delta } I_{1} (\epsilon).
\eeq
From this free energy, we can evaluate the impurity contribution to the specific heat 
\begin{align}
 C_\text{imp} &
 = - T \frac{\partial^2 f_\text{imp}}{\partial T^2}
 \nonumber \\[.5em]
 &
 =
 \begin{cases}\displaystyle
  \frac{ \lambda^{2} }{2} \pi^{1 + 2 \Delta} ( 2\Delta)^{2} (N^{2}-1) (N + k / 2 ) \left[ \frac{ 1 - 2 \Delta }{ 2 } \right] \frac{ \Gamma(1/2 - \Delta ) \Gamma(1/2) }{ \Gamma(1-\Delta ) } \, T^{2 \Delta} & (k > N) \\[1em]
  \displaystyle
  \lambda^{2} \pi^{1 + 2 \Delta} ( N^{2}-1 )( N + k/2 ) ( 2\Delta )^{2} \, T \, {\rm{log}} \left( \frac{T_{\rm{K}}}{T} \right) & (k = N) \\[1em]
    \displaystyle
  2 \lambda^{2} \pi^{2} (N^{2}-1) (N + k/2) \frac{ 2 \Delta }{ 1 + 2 \Delta } \left( \frac{\beta_{\rm{K}}^{-2 \Delta+1} }{ 2 \Delta -1 } \right)T & (N > k > 1)
 \end{cases}
 \, ,
 \label{eq:C_imp}
\end{align}
where $\beta_{\rm{K}} = 1 / T_{\rm{K}}$.
Although the specific heat contains the unknown constant $\lambda^{2}$, we can read off the temperature dependences of the specific heat at low temperatures as\footnote{In Refs.~\onlinecite{Parcollet:1997PRL,Parcollet:1998PRB}, the authors show the low temperature dependences of $C_{\rm{imp}}$, as well as $\chi_{\rm{imp}}$, in the context of the large $N$ analysis without the explicit form (\ref{eq:C_imp}). As we will see in Sec.~\ref{sec:Wratio}, the explicit forms of $C_{\rm{imp}}$ and $\chi_{\rm{imp}}$ enable us to obtain the Wilson ratio, and it turns out that the Wilson ratio is non-universal for $N > k > 1$.}
\begin{align}
 C_\text{imp}
 & \propto
 \begin{cases}
  T^{2 \Delta} & (k>N) \\
  T \log (T_\text{K}/T) & (k=N) \\
  T & (N > k > 1)
 \end{cases}
 \label{eq:C_dep}
\end{align}
where the dimension $\Delta$ is given by \eqref{eq:conf_dim}.
This shows fractional power (logarithmic) temperature dependence for $k \ge N$, suggesting the non-Fermi liquid behavior, while we obtain the linear dependence for $N > k > 1$.
The latter behavior is actually the same as the single-channel system $k = 1$ \eqref{eq:C_FL}, which is described as the Fermi-liquid.
On the other hand, the analysis of $g$-factor performed in Sec.~\ref{sec:g} shows that the IR fixed point is of the non-Fermi liquid for $N  > k > 1$, which is consistent with the Bethe ansatz analysis~\cite{Jerez:1998PRB}.
Although the fixed point is classified as the non-Fermi liquid, the low temperature behavior is described as the Fermi-liquid for the cases $N > k > 1$.

We remark that since Eqs.~\eqref{eq:C_FL} and \eqref{eq:C_imp} have the same scaling behavior as $T / T_{\rm{K}}$ for $N > k > 1$, they contribute to the specific heat in the same order. Thus, a proper contribution to the specific heat should be a combination of these two terms
\begin{align}
 C_\text{imp}
 & =
 - \lambda_1 \frac{k(N^2 - 1)}{3} \pi^2 T
 + 2 \lambda^{2} \pi^{2} (N^{2}-1) (N + k/2) \frac{ 2 \Delta }{ 1 + 2 \Delta } \left( \frac{\beta_{\rm{K}}^{-2 \Delta+1} }{ 2 \Delta -1 } \right)T 
 \quad (N > k > 1)
 \, ,
 \label{eq:C_mixed}
\end{align}
which shows the linear temperature dependence.
The situation is summarized as follows.
Although the operator dimension itself of $\delta \mathcal{H}$ \eqref{eq:pert2} is lower than $\delta \mathcal{H}_1$ \eqref{eq:pert1}, the perturbation starts from the second order of $\delta \mathcal{H}$ while it starts from the first order of $\delta \mathcal{H}_1$. Consequently, these two contributions to the free energy are compatible with each other
, and thus the contributions of the operators of the local Fermi liquid type and the non-Fermi liquid type are mixed up.
We call this peculiar behavior \textit{the Fermi/non-Fermi mixing}.
The combined expression of the specific heat (\ref{eq:C_mixed}) will play an important role when we discuss the Wilson ratio for $N > k > 1$ in Sec.~\ref{sec:Wratio}.

\subsubsection{Susceptibility}
\label{sec:chi}

Next we compute the low temperature behavior of ``$\SU(N)$ susceptibility'' from the two-point function $\langle \mathcal{J}^a(\tau_1,x_1) \mathcal{J}^a(\tau_2,x_2) \rangle$ by taking into account the external source.
Here we evaluate the impurity contribution to the susceptibility which is given by
 \beq
 \chi_{\rm{imp}}
 &=& - \left. \frac{ \partial^{2} f_{\rm{imp}} }{ \partial h^{2} } \right|_{h=0}, 
 \label{def_suscpt}
 \eeq
 with 
\beq
{\rm{e}}^{ - \beta f_{\rm{imp}} }
&=& Z_{0} \left\langle {\rm{exp}} \left\{ \int^{\beta/2}_{-\beta/2} d \tau \left[ \lambda \mathcal{J}^{a}_{-1} \phi^{a} ( \tau, 0 ) + \frac{ h }{ 2 \pi } \int^{\infty}_{- \infty} dx \mathcal{J}^{3} ( \tau, x ) \right]
\right\} \right\rangle
\eeq
Expanding the partition function with respect to both $\lambda$ and $h$, we find
\beq
{\rm{e}}^{ - \beta f_{\rm{imp}} }
&=& 1 + \frac{1}{2!} \left( \frac{ h }{ 2\pi } \right)^{2} \int d \tau_{1} d x_{1} \int d \tau_{2} d x_{2} 
\left\langle \mathcal{J}^{3}( \tau_{1}, x_{1} ) \mathcal{J}^{3} ( \tau_{2}, x_{2} ) \right\rangle \nonumber \\
&& + \frac{ \lambda^{2} }{ 2 ! } \int d \tau_{3} d \tau_{4} \left\langle \mathcal{J}^{a}_{-1} \phi^{a} ( \tau_{3}, 0 ) \mathcal{J}^{b}_{-1} \phi^{b} ( \tau_{4}, 0 ) \right\rangle \nonumber \\
&& + \frac{1}{2!} \left( \frac{ h }{ 2\pi } \right)^{2} \frac{ \lambda^{2} }{ 2! } \int d \tau_{1} d x_{1} \int d \tau_{2} dx_{2} \int d\tau_{3} d \tau_{4}
\left\langle \mathcal{J}^{3}( \tau_{1}, x_{1} ) \mathcal{J}^{3} ( \tau_{2}, x_{2} ) \mathcal{J}^{a}_{-1} \phi^{a} ( \tau_{3}, 0 ) \mathcal{J}^{b}_{-1} \phi^{b} ( \tau_{4}, 0 ) \right\rangle \nonumber \\
&& + \cdots
\eeq
Here we have omitted the upper and lower limits of the integrals, since the limits of each integral are clear without them.
Taking logarithm in both sides, we get
\beq
- \beta \delta f_{\rm{imp}}
&=&\frac{1}{2!} \left( \frac{ h }{ 2\pi } \right)^{2} \frac{ \lambda^{2} }{ 2! } \int d \tau_{1} d x_{1} \int d \tau_{2} dx_{2} \int d\tau_{3} d \tau_{4} \nonumber \\
&& \times \left\langle \mathcal{J}^{3}( \tau_{1}, x_{1} ) \mathcal{J}^{3} ( \tau_{2}, x_{2} ) \mathcal{J}^{a}_{-1} \phi^{a} ( \tau_{3}, 0 ) \mathcal{J}^{b}_{-1} \phi^{b} ( \tau_{4}, 0 ) \right\rangle_{\rm{conneted}} + \cdots.
\label{free_energy_lh}
\eeq
From Eqs. (\ref{def_suscpt}) and (\ref{free_energy_lh}), the susceptibility reads 
\beq
\chi_{\rm{imp}}
&=&\frac{ 1 }{ 2 \beta } \left( \frac{ \lambda }{ 2 \pi } \right)^{2} \int d \tau_{1} d x_{1} \int d \tau_{2} dx_{2} \int d\tau_{3} d \tau_{4}
\left\langle \mathcal{J}^{3}( \tau_{1}, x_{1} ) \mathcal{J}^{3} ( \tau_{2}, x_{2} ) \mathcal{J}^{a}_{-1} \phi^{a} ( \tau_{3}, 0 ) \mathcal{J}^{b}_{-1} \phi^{b} ( \tau_{4}, 0 ) \right\rangle_{\rm{conneted}},\nonumber \\
\eeq
up to the second order of $\lambda$.
Here, the connected correlation function can be expressed as \cite{Affleck:1990iv}
\begin{align}
&
 \left\langle \mathcal{J}^{3}( z_{1} ) \mathcal{J}^{3} ( z_{2} ) \mathcal{J}^{a}_{-1} \phi^{a} ( z_{3} ) \mathcal{J}^{b}_{-1} \phi^{b} ( z_{4} ) \right\rangle_{\rm{conneted}}
 \nonumber \\
 &= \left[ \frac{ 1 }{ \left( \frac{ \beta }{ \pi } {\rm{sin}} \frac{ \pi }{ \beta } ( z_{1} - z_{4} ) \right)^{2} } \frac{ 1 }{ \left( \frac{ \beta }{ \pi } {\rm{sin}} \frac{ \pi }{ \beta } ( z_{2} - z_{3} ) \right)^{2} }
\right. 
+ \left. \frac{1}{ \left( \frac{ \beta }{ \pi } {\rm{sin}} \frac{ \pi }{ \beta } ( z_{1} - z_{3} ) \right)^{2} } \frac{ 1 }{ \left( \frac{ \beta }{ \pi } {\rm{sin}} \frac{ \pi }{ \beta } ( z_{2} - z_{4} ) \right)^{2} } \right]
\nonumber \\
& \quad \times \frac{ ( N + k/2 )^{2} }{ \left( \frac{ \beta }{ \pi } {\rm{sin}} \frac{ \pi }{ \beta } ( z_{3} - z_{4} ) \right)^{2 \Delta } }
\end{align}
By using this correlation function, one can get the susceptibility as
\begin{align}
\chi_{\rm{imp}}
&= \frac{1}{2\beta} \left( \frac{ \lambda }{ 2\pi } \right)^{2} \int d \tau_{1} d x_{1} \int d \tau_{2} dx_{2} \int d\tau_{3} d \tau_{4} \nonumber \\
& \times \left[ \frac{1}{ \left( \frac{ \beta }{ \pi } {\rm{sin}} \frac{ \pi }{ \beta } ( (\tau_{1} - \tau_{4} ) + ix_{1} ) \right)^{2} }
\frac{1}{ \left( \frac{ \beta }{ \pi } {\rm{sin}} \frac{ \pi }{ \beta }( ( \tau_{2} - \tau_{3} ) + ix_{2} ) \right)^{2} } 
 + ( 1 \leftrightarrow 2 )  \right] \nonumber \\
& \times \frac{ ( N + k/2 )^{2} }{ \left| \frac{ \beta }{ \pi } {\rm{sin}} \frac{ \pi }{ \beta } ( \tau_{3} - \tau_{4} ) \right|^{2 \Delta } } 
\end{align}
Integrating $x_{1}$ and $x_{2}$, we find
\begin{align}
 \chi_\text{imp}
 & =
 \frac{1}{2\beta} \left( \frac{ \lambda }{ 2\pi } \right)^{2} 2 \left( \frac{ 2\pi }{ \beta } \right)^{2} \beta^{2} ( N + k/2 )^{2} \int d \tau_{3} d \tau_{4} \frac{1}{ \left| \frac{ \beta }{ \pi } {\rm{sin}} \frac{ \pi }{ \beta } ( \tau_{3} - \tau_{4} ) \right|^{2 \Delta } } \nonumber \\
 & =
 2 \lambda^{2} ( N + k/2 )^{2} \left( \frac{ \pi }{ \beta } \right)^{2 \Delta - 1 } I_{1} (\epsilon) \nonumber \\
 & =
 \begin{cases}\displaystyle
  \frac{\lambda^2}{2} \pi^{2\Delta -1 } (N + k/2 )^{2} (1-2\Delta) \frac{ \Gamma(1/2 - \Delta ) \Gamma(1/2) }{ \Gamma(1-\Delta ) } \, T^{2\Delta - 1} & (k > N) \\[1em] \displaystyle
  2 \lambda^2 (N + k/2 )^{2} \, \log \left(\frac{T_\text{K}}{T}\right) & (k = N) \\[1em] \displaystyle
  2 \lambda^{2} (N + k/2)^{2} \left( \frac{ \beta_{\rm{K}}^{-2 \Delta + 1} }{ 2 \Delta -1 } \right) & (N > k > 1)
 \end{cases}
 \, .
 \label{eq:chi_imp}
\end{align}
Now the scaling behavior of the susceptibility is summarized as
\begin{align}
 \chi_\text{imp}
 & =
 \begin{cases}
  T^{2\Delta - 1} & (k > N) \\
  \log(T_\text{K}/T) & (k = N) \\
  \text{const.} & (N > k > 1)
 \end{cases}
\end{align}
where the dimension $\Delta$ is given by \eqref{eq:conf_dim}.
We obtain fractional power (logarithmic) temperature dependence for $k \ge N$, while there is no temperature dependence for $N > k > 1$, which is the same as the single-channel system \eqref{eq:chi_FL}.
As discussed in Sec.~\ref{sec:C}, the low temperature behavior for $N > k > 1$ implies the Fermi/non-Fermi mixing behavior, which is the Fermi liquid-type excitation from the non-Fermi liquid IR fixed point, and consistent with the Bethe ansatz analysis~\cite{Jerez:1998PRB}.
Again, the susceptibility (\ref{eq:chi_imp}) for $N > k > 1$ is the same order of Eq. (\ref{eq:chi_FL}) with respect to $T_{\rm{K}}$.
Therefore, the proper expression of the susceptibility for $N > k > 1$ is the combination of these two contributions as
\begin{align}
 \chi_\text{imp}
 & =
 - \lambda_1 \frac{k(N+k)}{2}
 + 2 \lambda^{2} (N + k/2)^{2} \left( \frac{ \beta_{\rm{K}}^{-2 \Delta + 1} }{ 2 \Delta -1 } \right)
 \qquad (N > k > 1)
 \, .
 \label{eq:chi_mixed}
\end{align}
This shows the Fermi/non-Fermi mixing, the mixing of two contributions of Fermi/non-Fermi type operators, as in the specific heat.

\subsubsection{Wilson Ratio}
\label{sec:Wratio}

Now we can evaluate the Wilson ratio~\cite{Wilson:1974mb} from the impurity contributions obtained in the previous subsections.
For the single-channel system ($k=1$), it is given by \eqref{eq:Wratio1}.
For $k \ge N > 1$, the ratio is given by~\cite{Ludwig:1994nf}
\begin{align}
 R_\text{W}
 & =
 \frac{(N + k/2)(N + k)^2}{3N (N^2 - 1)}
 \qquad (k \ge N > 1)
 \, ,
\end{align}
which is a universal constant characterizing the IR fixed point, not depending on the unknown parameter $\lambda$ as well as $T_{\rm{K}}$.
For $N > k > 1$, on the other hand, we have to take into account the two impurity contributions as shown in \eqref{eq:C_mixed} and \eqref{eq:chi_mixed}.
The Wilson ratio is then given by
\begin{align}
 R_\text{W}
 & =
 \frac{2\pi^2}{3} N T \,
 \frac{\chi_\text{imp}}{C_\text{imp}}
 \nonumber \\
 & =
 \frac{(N+k/2)(N+k/3)}{N^2-1}
 \frac{\displaystyle \gamma - \frac{k(N+k)}{(N+k/2)^2}}
      {\displaystyle \gamma - \frac{k(N+k/3)}{N (N+k/2)}}
 \qquad
 (N > k > 1)
 \, .
 \label{WilsonRatio2}
 \end{align}
where the dimensionless constant is defined as
\begin{align}
 \gamma & =
 4 \frac{\lambda^{2}}{\lambda_1} T_\text{K}^{2\Delta-1} 
 \, .
\end{align}
In this case the Wilson ratio depends on two coupling constants $(\lambda_1, \lambda)$ as well as $T_{\rm{K}}$ through the dimensionless constant $\gamma$, which depend on the details of the system. 
This indicates that the Wilson ratio no longer plays a role as the universal quantity characterizing the IR fixed point for $N > k > 1$.
We remark that the Wilson ratio \eqref{WilsonRatio2} reproduces the result for the local Fermi liquid when $\gamma=0$ (turning off the non-Fermi liquid type operator $\lambda = 0$) and for the non-Fermi liquid when $\gamma=\infty$ (turning off the Fermi liquid type operator $\lambda_1=0$).
Let us comment on another approach to the Kondo model in the region $N > k > 1$.
The Wilson ratio is numerically evaluated using Bethe ansatz~\cite{Jerez:1998PRB}, and their result in the case $N > k > 1$ corresponds to that of CFT without the non-Fermi liquid type operator \eqref{eq:pert2}, namely $\lambda= 0$, or equivalently $\gamma = 0$. We cannot figure out the reason for this behavior in the CFT approach because the leading irrelevant operator \eqref{eq:pert2} is always possible when $k > 1$ irrespective of the relation between $N$ and $k$, and thus there is no reason to suppress the the non-Fermi liquid type operator in the case of $N > k > 1$ from the view point of the symmetry.
Since other observables such as $g$-factor, the low $T$ behaviors of the specific heart, and susceptibility are consistent with the Bethe ansatz, it is interesting to explore the discrepancy between the Wilson ratios form CFT and the Bethe ansatz in the case of $N > k > 1$ by using other non-perturbative approaches including the numerical renormalization group, and from future experiments of multi-channel Kondo effect.

In the large $N$ limit, we find that the Wilson ratio approaches to unity
\beq
R_{\rm{W}}
\to 1  \qquad (N \to \infty).
\eeq
Therefore, the Wilson ratio again becomes an universal quantity, which is consistent with the single-channel system \eqref{eq:Wratio1} in the limit $N \to \infty$.
In this case, the $g$-factor also behaves as the Fermi liquid.

We expect that the non-universal property of the Wilson ratio (\ref{WilsonRatio2}) for $N > k > 1$ can be used as an experimental signal of the Fermi/non-Fermi mixing. Because the the low temperature scaling behavior of the single-channel system described as the Fermi liquid is the same as that of the multi-channel system with $N > k > 1$, it is difficult to distinguish these two systems just by observing the low temperature dependences of the observables such as the specific heat and the susceptibility.
However, the Wilson ratio for $N > k > 1$ deviates from the universal value of the Wilson ratio with $k=1$. The deviation provides information of the Fermi/non-Fermi mixing in the multi-channel systems with $N > k > 1$. This enables us to experimentally observe the Fermi/non-Fermi mixing, and gives some insights into the discrepancy of the Wilson ratio between CFT approach and the Bethe ansatz in the case of $N > k > 1$.


\section{Discussion}
\label{sec:discussion}

In this paper, we have studied the multi-channel SU($N$) Kondo system by using the conformal field theory approach.
We have computed the $g$-factor, associated with the impurity entropy, to characterize the IR fixed point of $k$-channel SU($N$) Kondo system.
From the $g$-factor analysis, we have shown that the zero temperature IR fixed point is described as the Fermi liquid for single-channel system $k=1$, while as the non-Fermi liquid for multi-channel systems $k>1$.
We have then investigated the low temperature behavior of the specific heat and susceptibility.
The scaling behavior depends on 
the fundamental parameters $(N, k)$ of the multi-channel Kondo system.
For the cases $N \le k$, we have observed the fractional power (logarithmic) temperature scaling, implying the non-Fermi liquid behavior.
On the other hand, we have observed the integer power scaling, which is the Fermi liquid behavior, even for the multi-channel system with $N > k > 1$.
This means that the Fermi liquid-type low temperature excitation is observed even if the corresponding IR fixed point is of the non-Fermi liquid for the multi-channel Kondo system with $N > k > 1$.
In this case we have to take into account two kinds of irrelevant operators to perform the perturbation analysis since their contributions are compatible.
As a result, the Wilson ratio no longer plays a role of the universal quantity, which depends on the microscopic details of the system. 
We have shown the mixing of two contributions from the Fermi/non-Fermi type operators, which we call the Fermi/non-Fermi mixing, based on the CFT analysis.
However the corresponding microscopic electronic state cannot be clarified by the current CFT method.
In order to elucidate the electronic state which causes the mixing behaviors,
it would be important to study the spectrum structure of the electron with $N > k > 1$. This will be our future work beyond this study.

The Fermi/non-Fermi mixing discussed in this paper can be experimentally observed in the multi-channel Kondo system with $N > k > 1$.
There are several possibilities to realize such a system in experiments.
The first possibility is to use the quantum dot exhibiting the SU(4) Kondo effect~\cite{Sasaki:2004PRL,Jarillo-Herrero:2005Nat}.
In particular, the two-channel SU(4) system satisfies the condition $N > k >1$ with $(N,k) = (4,2)$.
It is a strong candidate to realize the Fermi/non-Fermi mixing.
The second is to use the ultracold atom showing the SU(3) Kondo effect due to the orbital degeneracy~\cite{Nishida:2013PRL,Nishida:2016PRA}.
If it is possible to realize the two-channel system in this case, the system shows the Fermi/non-Fermi mixing since $(N,k) = (3,2)$.
Recently, the Kondo effects induced by the strong interaction are proposed in QCD and nuclear physics \cite{Yasui:2013xr,Hattori:2015hka,Ozaki:2015sya,Yasui:2016ngy,Yasui:2016svc,Yasui:2016hlz,Yasui:2016yet, Kanazawa:2016ihl}. Among them, the QCD Kondo effect \cite{Hattori:2015hka,Ozaki:2015sya} induced by the color SU(3) interaction (or more generally the color SU($N_{c}$) interaction) is an important possibility to realize the Fermi/non-Fermi mixing. Because the QCD naturally has the flavor degrees of freedom, the QCD Kondo effect corresponds to the multi-channel Kondo effect. Depending on the numbers of the color and the flavor, the Fermi/non-Fermi mixing is realized in QCD Kondo effect in the low temperature region below the Kondo temperature~\cite{Kimura:2016inp}.
It is also interesting if the Fermi/non-Fermi mixing is observed in other non-perturbative methods such as the numerical renormalization group and so on. 

\section*{Acknowledgements}

We would like to thank A.~Furusaki, S.~C.~Furuya, and R.~Yoshii for useful discussion.
The work of TK and SO was supported in part by MEXT-Supported Program for the Strategic Research Foundation at Private Universities ``Topological Science'' (No.~S1511006).
TK was also supported in part by Keio Gijuku Academic Development Funds, JSPS Grant-in-Aid for Scientific Research (No.~JP17K18090), and JSPS Grant-in-Aid for Scientific Research on Innovative Areas ``Topological Materials Science'' (No.~JP15H05855).

\appendix
\section{Quantum dimension}
\label{sec:Schur}

The quantum dimension is obtained, in practice, using the Schur polynomial.
The Schur polynomial, which is an $N$-variable symmetric polynomial for $\SU(N)$ theory, is defined~\cite{Macdonald:1997}
\begin{align}
 s_\lambda(x_1,\ldots,x_N) & =
 \frac{1}{\Delta(x)} \det_{1 \le i, j \le N} x_i^{\lambda_j + N - j}
\end{align}
where $\lambda$ is the partition and the Vandermonde determinant defined
\begin{align}
 \Delta(x)
 = \det_{1 \le i, j \le N} x_i^{N-j}
 = \prod_{i<j}^N (x_i - x_j)
 \, .
\end{align}
The quantum dimension is then given by the principal specialization of the Schur polynomial,
\begin{align}
 \operatorname{dim}_q R & = s_R(q^\rho)
\end{align}
where the $\SU(N)$ Weyl vector
\begin{align}
 \rho = \left\{\frac{N+1}{2} - i\right\}_{i=1,\ldots,N}
 \, .
 \label{eq:Weyl_vec}
\end{align}
We obtain a formula \eqref{eq:q-dim} with this definition, which computes the quantum dimension.
The formula \eqref{eq:q-dim} depends only on differences of entries $R_i - R_j$, which leads to the invariance under the constant shift $R_i \to R_i + c$ for $\forall i$.
Thus we can fix the last element as $R_N = 0$ without loss of generality.

For example, the $\SU(2)$ representation is classified into $s$-spin representation $R_s$, which corresponds to $(R_1, R_2) = (2s, 0)$, and the $q$-parameter is given by $q = \exp \left( 2 \pi i /(2 + k)\right)$.
The quantum dimension yields
\begin{align}
 \operatorname{dim}_q R_s
 & = [2s+1]_q
 = \frac{\displaystyle \sin \left( \frac{2s+1}{2+k}\pi \right)}
        {\displaystyle \sin \left( \frac{\pi}{2+k}\right)}
 \, .
\end{align}
In the limit $q \to 1$, this reproduces the dimension of $s$-spin representation, $\operatorname{dim} R_s = 2s+1$.

We then show an explicit formula for $\SU(3)$ theory.
Parametrizing $(R_1, R_2, R_3) = (r,s,0)$ with $r \ge s$, the quantum dimension is given by
\begin{align}
 \operatorname{dim}_q R
 & =
 \frac{\left( q^{\frac{1}{2}(r+2)} - q^{-\frac{1}{2}(r+2)} \right)}
      {\Big( q - q^{-1} \Big)}
 \frac{\left( q^{\frac{1}{2}(s+1)} - q^{-\frac{1}{2}(s+1)} \right)}
      {\left( q^{\frac{1}{2}} - q^{-\frac{1}{2}} \right)}
 \frac{\left( q^{\frac{1}{2}(r-s+1)} - q^{-\frac{1}{2}(r-s+1)} \right)}
      {\left( q^{\frac{1}{2}} - q^{-\frac{1}{2}} \right)} 
 \nonumber \\
 & =
 \frac{\displaystyle
       \sin \left( \frac{r+2}{3+k}\pi \right)}
      {\displaystyle
       \sin \left( \frac{2\pi}{3+k} \right)}
 \frac{\displaystyle
       \sin \left( \frac{s+1}{3+k}\pi \right)}
      {\displaystyle
       \sin \left( \frac{\pi}{3+k} \right)}
 \frac{\displaystyle
       \sin \left( \frac{r-s+1}{3+k}\pi \right)}
      {\displaystyle
       \sin \left( \frac{\pi}{3+k} \right)}
\end{align}
where the $q$-parameter is taken to be $q = \exp\left(2\pi i/(3+k)\right)$, and $k$ is the level.

\bibliography{qcdkondo}

\begin{thebibliography}{39}%
\makeatletter
\providecommand \@ifxundefined [1]{%
 \@ifx{#1\undefined}
}%
\providecommand \@ifnum [1]{%
 \ifnum #1\expandafter \@firstoftwo
 \else \expandafter \@secondoftwo
 \fi
}%
\providecommand \@ifx [1]{%
 \ifx #1\expandafter \@firstoftwo
 \else \expandafter \@secondoftwo
 \fi
}%
\providecommand \natexlab [1]{#1}%
\providecommand \enquote  [1]{``#1''}%
\providecommand \bibnamefont  [1]{#1}%
\providecommand \bibfnamefont [1]{#1}%
\providecommand \citenamefont [1]{#1}%
\providecommand \href@noop [0]{\@secondoftwo}%
\providecommand \href [0]{\begingroup \@sanitize@url \@href}%
\providecommand \@href[1]{\@@startlink{#1}\@@href}%
\providecommand \@@href[1]{\endgroup#1\@@endlink}%
\providecommand \@sanitize@url [0]{\catcode `\\12\catcode `\$12\catcode
  `\&12\catcode `\#12\catcode `\^12\catcode `\_12\catcode `\%12\relax}%
\providecommand \@@startlink[1]{}%
\providecommand \@@endlink[0]{}%
\providecommand \url  [0]{\begingroup\@sanitize@url \@url }%
\providecommand \@url [1]{\endgroup\@href {#1}{\urlprefix }}%
\providecommand \urlprefix  [0]{URL }%
\providecommand \Eprint [0]{\href }%
\providecommand \doibase [0]{http://dx.doi.org/}%
\providecommand \selectlanguage [0]{\@gobble}%
\providecommand \bibinfo  [0]{\@secondoftwo}%
\providecommand \bibfield  [0]{\@secondoftwo}%
\providecommand \translation [1]{[#1]}%
\providecommand \BibitemOpen [0]{}%
\providecommand \bibitemStop [0]{}%
\providecommand \bibitemNoStop [0]{.\EOS\space}%
\providecommand \EOS [0]{\spacefactor3000\relax}%
\providecommand \BibitemShut  [1]{\csname bibitem#1\endcsname}%
\let\auto@bib@innerbib\@empty
\bibitem [{\citenamefont {Kondo}(1964)}]{Kondo:1964nea}%
  \BibitemOpen
  \bibfield  {author} {\bibinfo {author} {\bibfnamefont {J.}~\bibnamefont
  {Kondo}},\ }\href {\doibase 10.1143/PTP.32.37} {\bibfield  {journal}
  {\bibinfo  {journal} {Prog. Theor. Phys.}\ }\textbf {\bibinfo {volume}
  {32}},\ \bibinfo {pages} {37} (\bibinfo {year} {1964})}\BibitemShut {NoStop}%
\bibitem [{\citenamefont {Wilson}(1975)}]{Wilson:1974mb}%
  \BibitemOpen
  \bibfield  {author} {\bibinfo {author} {\bibfnamefont {K.~G.}\ \bibnamefont
  {Wilson}},\ }\href {\doibase 10.1103/RevModPhys.47.773} {\bibfield  {journal}
  {\bibinfo  {journal} {Rev. Mod. Phys.}\ }\textbf {\bibinfo {volume} {47}},\
  \bibinfo {pages} {773} (\bibinfo {year} {1975})}\BibitemShut {NoStop}%
\bibitem [{\citenamefont {Andrei}(1980)}]{Andrei:1980fv}%
  \BibitemOpen
  \bibfield  {author} {\bibinfo {author} {\bibfnamefont {N.}~\bibnamefont
  {Andrei}},\ }\href {\doibase 10.1103/PhysRevLett.45.379} {\bibfield
  {journal} {\bibinfo  {journal} {Phys. Rev. Lett.}\ }\textbf {\bibinfo
  {volume} {45}},\ \bibinfo {pages} {379} (\bibinfo {year} {1980})}\BibitemShut
  {NoStop}%
\bibitem [{\citenamefont {Wiegmann}(1980)}]{Wiegmann:1980JETP}%
  \BibitemOpen
  \bibfield  {author} {\bibinfo {author} {\bibfnamefont {P.}~\bibnamefont
  {Wiegmann}},\ }\href
  {http://www.jetpletters.ac.ru/ps/1353/article_20434.shtml} {\bibfield
  {journal} {\bibinfo  {journal} {JETP Letters}\ }\textbf {\bibinfo {volume}
  {31}},\ \bibinfo {pages} {364} (\bibinfo {year} {1980})}\BibitemShut
  {NoStop}%
\bibitem [{\citenamefont {Coleman}(1983)}]{Coleman:1983PRB}%
  \BibitemOpen
  \bibfield  {author} {\bibinfo {author} {\bibfnamefont {P.}~\bibnamefont
  {Coleman}},\ }\href {\doibase 10.1103/PhysRevB.28.5255} {\bibfield  {journal}
  {\bibinfo  {journal} {Phys. Rev.}\ }\textbf {\bibinfo {volume} {B28}},\
  \bibinfo {pages} {5255} (\bibinfo {year} {1983})}\BibitemShut {NoStop}%
\bibitem [{\citenamefont {Read}\ and\ \citenamefont
  {Newns}(1983)}]{Read:1983JPC}%
  \BibitemOpen
  \bibfield  {author} {\bibinfo {author} {\bibfnamefont {N.}~\bibnamefont
  {Read}}\ and\ \bibinfo {author} {\bibfnamefont {D.~M.}\ \bibnamefont
  {Newns}},\ }\href {\doibase 10.1088/0022-3719/16/17/014} {\bibfield
  {journal} {\bibinfo  {journal} {J. Phys. C: Solid State Physics}\ }\textbf
  {\bibinfo {volume} {16}},\ \bibinfo {pages} {3273} (\bibinfo {year}
  {1983})}\BibitemShut {NoStop}%
\bibitem [{\citenamefont {Coleman}(1987)}]{Coleman:1987PRB}%
  \BibitemOpen
  \bibfield  {author} {\bibinfo {author} {\bibfnamefont {P.}~\bibnamefont
  {Coleman}},\ }\href {\doibase 10.1103/PhysRevB.35.5072} {\bibfield  {journal}
  {\bibinfo  {journal} {Phys. Rev.}\ }\textbf {\bibinfo {volume} {B35}},\
  \bibinfo {pages} {5072} (\bibinfo {year} {1987})}\BibitemShut {NoStop}%
\bibitem [{\citenamefont {Parcollet}\ and\ \citenamefont
  {Georges}(1997)}]{Parcollet:1997PRL}%
  \BibitemOpen
  \bibfield  {author} {\bibinfo {author} {\bibfnamefont {O.}~\bibnamefont
  {Parcollet}}\ and\ \bibinfo {author} {\bibfnamefont {A.}~\bibnamefont
  {Georges}},\ }\href {\doibase 10.1103/PhysRevLett.79.4665} {\bibfield
  {journal} {\bibinfo  {journal} {Phys. Rev. Lett.}\ }\textbf {\bibinfo
  {volume} {79}},\ \bibinfo {pages} {4665} (\bibinfo {year} {1997})},\ \Eprint
  {http://arxiv.org/abs/cond-mat/9707337} {cond-mat/9707337 [cond-mat.str-el]}
  \BibitemShut {NoStop}%
\bibitem [{\citenamefont {Parcollet}\ \emph {et~al.}(1998)\citenamefont
  {Parcollet}, \citenamefont {Georges}, \citenamefont {Kotliar},\ and\
  \citenamefont {Sengupta}}]{Parcollet:1998PRB}%
  \BibitemOpen
  \bibfield  {author} {\bibinfo {author} {\bibfnamefont {O.}~\bibnamefont
  {Parcollet}}, \bibinfo {author} {\bibfnamefont {A.}~\bibnamefont {Georges}},
  \bibinfo {author} {\bibfnamefont {G.}~\bibnamefont {Kotliar}}, \ and\
  \bibinfo {author} {\bibfnamefont {A.}~\bibnamefont {Sengupta}},\ }\href
  {\doibase 10.1103/PhysRevB.58.3794} {\bibfield  {journal} {\bibinfo
  {journal} {Phys. Rev.}\ }\textbf {\bibinfo {volume} {B58}},\ \bibinfo {pages}
  {3794} (\bibinfo {year} {1998})},\ \Eprint
  {http://arxiv.org/abs/cond-mat/9711192} {cond-mat/9711192 [cond-mat.str-el]}
  \BibitemShut {NoStop}%
\bibitem [{\citenamefont {Affleck}(1990)}]{Affleck:1990zd}%
  \BibitemOpen
  \bibfield  {author} {\bibinfo {author} {\bibfnamefont {I.}~\bibnamefont
  {Affleck}},\ }\href {\doibase 10.1016/0550-3213(90)90440-O} {\bibfield
  {journal} {\bibinfo  {journal} {Nucl. Phys.}\ }\textbf {\bibinfo {volume}
  {B336}},\ \bibinfo {pages} {517} (\bibinfo {year} {1990})}\BibitemShut
  {NoStop}%
\bibitem [{\citenamefont {Affleck}\ and\ \citenamefont
  {Ludwig}(1991{\natexlab{a}})}]{Affleck:1991tk}%
  \BibitemOpen
  \bibfield  {author} {\bibinfo {author} {\bibfnamefont {I.}~\bibnamefont
  {Affleck}}\ and\ \bibinfo {author} {\bibfnamefont {A.~W.~W.}\ \bibnamefont
  {Ludwig}},\ }\href {\doibase 10.1103/PhysRevLett.67.161} {\bibfield
  {journal} {\bibinfo  {journal} {Phys. Rev. Lett.}\ }\textbf {\bibinfo
  {volume} {67}},\ \bibinfo {pages} {161} (\bibinfo {year}
  {1991}{\natexlab{a}})}\BibitemShut {NoStop}%
\bibitem [{\citenamefont {Affleck}\ and\ \citenamefont
  {Ludwig}(1991{\natexlab{b}})}]{Affleck:1990by}%
  \BibitemOpen
  \bibfield  {author} {\bibinfo {author} {\bibfnamefont {I.}~\bibnamefont
  {Affleck}}\ and\ \bibinfo {author} {\bibfnamefont {A.~W.~W.}\ \bibnamefont
  {Ludwig}},\ }\href {\doibase 10.1016/0550-3213(91)90109-B} {\bibfield
  {journal} {\bibinfo  {journal} {Nucl. Phys.}\ }\textbf {\bibinfo {volume}
  {B352}},\ \bibinfo {pages} {849} (\bibinfo {year}
  {1991}{\natexlab{b}})}\BibitemShut {NoStop}%
\bibitem [{\citenamefont {Affleck}\ and\ \citenamefont
  {Ludwig}(1991{\natexlab{c}})}]{Affleck:1990iv}%
  \BibitemOpen
  \bibfield  {author} {\bibinfo {author} {\bibfnamefont {I.}~\bibnamefont
  {Affleck}}\ and\ \bibinfo {author} {\bibfnamefont {A.~W.~W.}\ \bibnamefont
  {Ludwig}},\ }\href {\doibase 10.1016/0550-3213(91)90419-X} {\bibfield
  {journal} {\bibinfo  {journal} {Nucl. Phys.}\ }\textbf {\bibinfo {volume}
  {B360}},\ \bibinfo {pages} {641} (\bibinfo {year}
  {1991}{\natexlab{c}})}\BibitemShut {NoStop}%
\bibitem [{\citenamefont {Affleck}\ and\ \citenamefont
  {Ludwig}(1993)}]{Affleck:1992ng}%
  \BibitemOpen
  \bibfield  {author} {\bibinfo {author} {\bibfnamefont {I.}~\bibnamefont
  {Affleck}}\ and\ \bibinfo {author} {\bibfnamefont {A.~W.~W.}\ \bibnamefont
  {Ludwig}},\ }\href {\doibase 10.1103/PhysRevB.48.7297} {\bibfield  {journal}
  {\bibinfo  {journal} {Phys. Rev.}\ }\textbf {\bibinfo {volume} {B48}},\
  \bibinfo {pages} {7297} (\bibinfo {year} {1993})}\BibitemShut {NoStop}%
\bibitem [{\citenamefont {Ludwig}\ and\ \citenamefont
  {Affleck}(1994)}]{Ludwig:1994nf}%
  \BibitemOpen
  \bibfield  {author} {\bibinfo {author} {\bibfnamefont {A.~W.~W.}\
  \bibnamefont {Ludwig}}\ and\ \bibinfo {author} {\bibfnamefont
  {I.}~\bibnamefont {Affleck}},\ }\href {\doibase 10.1016/0550-3213(94)90365-4}
  {\bibfield  {journal} {\bibinfo  {journal} {Nucl. Phys.}\ }\textbf {\bibinfo
  {volume} {B428}},\ \bibinfo {pages} {545} (\bibinfo {year}
  {1994})}\BibitemShut {NoStop}%
\bibitem [{\citenamefont {Affleck}(1995)}]{Affleck:1995ge}%
  \BibitemOpen
  \bibfield  {author} {\bibinfo {author} {\bibfnamefont {I.}~\bibnamefont
  {Affleck}},\ }\href {http://www.actaphys.uj.edu.pl/vol26/abs/v26p1869}
  {\bibfield  {journal} {\bibinfo  {journal} {Acta Phys. Polon.}\ }\textbf
  {\bibinfo {volume} {B26}},\ \bibinfo {pages} {1869} (\bibinfo {year}
  {1995})},\ \Eprint {http://arxiv.org/abs/cond-mat/9512099}
  {arXiv:cond-mat/9512099 [cond-mat]} \BibitemShut {NoStop}%
\bibitem [{\citenamefont {Jerez}\ \emph {et~al.}(1998)\citenamefont {Jerez},
  \citenamefont {Andrei},\ and\ \citenamefont {Zar{\'{a}}nd}}]{Jerez:1998PRB}%
  \BibitemOpen
  \bibfield  {author} {\bibinfo {author} {\bibfnamefont {A.}~\bibnamefont
  {Jerez}}, \bibinfo {author} {\bibfnamefont {N.}~\bibnamefont {Andrei}}, \
  and\ \bibinfo {author} {\bibfnamefont {G.}~\bibnamefont {Zar{\'{a}}nd}},\
  }\href {\doibase 10.1103/PhysRevB.58.3814} {\bibfield  {journal} {\bibinfo
  {journal} {Phys. Rev.}\ }\textbf {\bibinfo {volume} {B58}},\ \bibinfo {pages}
  {3814} (\bibinfo {year} {1998})},\ \Eprint
  {http://arxiv.org/abs/cond-mat/9803137} {arXiv:cond-mat/9803137
  [cond-mat.str-el]} \BibitemShut {NoStop}%
\bibitem [{Note1()}]{Note1}%
  \BibitemOpen
  \bibinfo {note} {The modular $S$-matrix formula for $\protect \mathrm
  {SU}(N)$ theory is given by \begin {align*} S_{\mu \nu } & = q^{-\protect
  \frac {1}{N} c(\mu ) c(\nu )} \protect \qopname \relax m{det}_{1 \le i, j \le
  N} q^{(\mu _i + N - i)(\nu _j + N - j)} \end {align*} where $\mu $ and $\nu $
  are partitions characterizing the $\protect \mathrm {SU}(N)$ representation,
  e.g. $\mu = (\mu _1, \mu _2, \protect \ldots , \mu _N) \in \protect \mathbb
  {Z}_{\ge 0}^N$, obeying $\mu _1 \ge \mu _2 \ge \protect \cdots \ge \mu _N \ge
  0$. The parameter $q$ is defined as $q = \protect \qopname \relax o{exp}\left
  ( 2 \pi i /(N+k)\right )$ for $\protect \mathrm {SU}(N)_k$ theory. See, for
  example, Ref.~\protect \rev@citealpnum {DiFrancesco:1997nk}. In this paper,
  we use the same notation for the partition and representation as long as no
  confusion. The constant $c(\mu )$ is defined as $\displaystyle c(\mu ) =
  \DOTSB \sum@ \slimits@ _{i=1}^N (\mu _i + N - i)$.}\BibitemShut {Stop}%
\bibitem [{\citenamefont {Witten}(1989)}]{Witten:1988hf}%
  \BibitemOpen
  \bibfield  {author} {\bibinfo {author} {\bibfnamefont {E.}~\bibnamefont
  {Witten}},\ }\href {\doibase 10.1007/BF01217730} {\bibfield  {journal}
  {\bibinfo  {journal} {Commun. Math. Phys.}\ }\textbf {\bibinfo {volume}
  {121}},\ \bibinfo {pages} {351} (\bibinfo {year} {1989})}\BibitemShut
  {NoStop}%
\bibitem [{\citenamefont {Di~Francesco}\ \emph {et~al.}(1997)\citenamefont
  {Di~Francesco}, \citenamefont {Mathieu},\ and\ \citenamefont
  {S\'en\'echal}}]{DiFrancesco:1997nk}%
  \BibitemOpen
  \bibfield  {author} {\bibinfo {author} {\bibfnamefont {P.}~\bibnamefont
  {Di~Francesco}}, \bibinfo {author} {\bibfnamefont {P.}~\bibnamefont
  {Mathieu}}, \ and\ \bibinfo {author} {\bibfnamefont {D.}~\bibnamefont
  {S\'en\'echal}},\ }\href {\doibase 10.1007/978-1-4612-2256-9} {\emph
  {\bibinfo {title} {{Conformal Field Theory}}}},\ Graduate Texts in
  Contemporary Physics\ (\bibinfo  {publisher} {Springer-Verlag},\ \bibinfo
  {address} {New York},\ \bibinfo {year} {1997})\BibitemShut {NoStop}%
\bibitem [{\citenamefont {Bl\"ote}\ \emph {et~al.}(1986)\citenamefont
  {Bl\"ote}, \citenamefont {Cardy},\ and\ \citenamefont
  {Nightingale}}]{Bloete:1986qm}%
  \BibitemOpen
  \bibfield  {author} {\bibinfo {author} {\bibfnamefont {H.~W.~J.}\
  \bibnamefont {Bl\"ote}}, \bibinfo {author} {\bibfnamefont {J.~L.}\
  \bibnamefont {Cardy}}, \ and\ \bibinfo {author} {\bibfnamefont {M.~P.}\
  \bibnamefont {Nightingale}},\ }\href {\doibase 10.1103/PhysRevLett.56.742}
  {\bibfield  {journal} {\bibinfo  {journal} {Phys. Rev. Lett.}\ }\textbf
  {\bibinfo {volume} {56}},\ \bibinfo {pages} {742} (\bibinfo {year}
  {1986})}\BibitemShut {NoStop}%
\bibitem [{\citenamefont {Affleck}(1986{\natexlab{a}})}]{Affleck:1986bv}%
  \BibitemOpen
  \bibfield  {author} {\bibinfo {author} {\bibfnamefont {I.}~\bibnamefont
  {Affleck}},\ }\href {\doibase 10.1103/PhysRevLett.56.746} {\bibfield
  {journal} {\bibinfo  {journal} {Phys. Rev. Lett.}\ }\textbf {\bibinfo
  {volume} {56}},\ \bibinfo {pages} {746} (\bibinfo {year}
  {1986}{\natexlab{a}})}\BibitemShut {NoStop}%
\bibitem [{\citenamefont {Affleck}(1986{\natexlab{b}})}]{Affleck:1986sc}%
  \BibitemOpen
  \bibfield  {author} {\bibinfo {author} {\bibfnamefont {I.}~\bibnamefont
  {Affleck}},\ }\href {\doibase 10.1103/PhysRevLett.56.2763} {\bibfield
  {journal} {\bibinfo  {journal} {Phys. Rev. Lett.}\ }\textbf {\bibinfo
  {volume} {56}},\ \bibinfo {pages} {2763} (\bibinfo {year}
  {1986}{\natexlab{b}})}\BibitemShut {NoStop}%
\bibitem [{Note2()}]{Note2}%
  \BibitemOpen
  \bibinfo {note} {For $k=1$, one cannot have the operator $\protect \mathcal
  {O}= \protect \mathcal {J}_{-1}^a \phi ^a$ since the adjoint primary,
  corresponding to $R=(2,1,\protect \ldots ,1,0)$, is not generated by the
  fusion process.}\BibitemShut {Stop}%
\bibitem [{Note3()}]{Note3}%
  \BibitemOpen
  \bibinfo {note} {In Refs.~\protect \rev@citealpnum
  {Parcollet:1997PRL,Parcollet:1998PRB}, the authors show the low temperature
  dependences of $C_{\protect \rm {imp}}$, as well as $\chi _{\protect \rm
  {imp}}$, in the context of the large $N$ analysis without the explicit form
  (\ref {eq:C_imp}). As we will see in Sec.~\ref {sec:Wratio}, the explicit
  forms of $C_{\protect \rm {imp}}$ and $\chi _{\protect \rm {imp}}$ enable us
  to obtain the Wilson ratio, and it turns out that the Wilson ratio is
  non-universal for $N > k > 1$.}\BibitemShut {Stop}%
\bibitem [{\citenamefont {Sasaki}\ \emph {et~al.}(2004)\citenamefont {Sasaki},
  \citenamefont {Amaha}, \citenamefont {Asakawa}, \citenamefont {Eto},\ and\
  \citenamefont {Tarucha}}]{Sasaki:2004PRL}%
  \BibitemOpen
  \bibfield  {author} {\bibinfo {author} {\bibfnamefont {S.}~\bibnamefont
  {Sasaki}}, \bibinfo {author} {\bibfnamefont {S.}~\bibnamefont {Amaha}},
  \bibinfo {author} {\bibfnamefont {N.}~\bibnamefont {Asakawa}}, \bibinfo
  {author} {\bibfnamefont {M.}~\bibnamefont {Eto}}, \ and\ \bibinfo {author}
  {\bibfnamefont {S.}~\bibnamefont {Tarucha}},\ }\href {\doibase
  10.1103/PhysRevLett.93.017205} {\bibfield  {journal} {\bibinfo  {journal}
  {Phys. Rev. Lett.}\ }\textbf {\bibinfo {volume} {93}},\ \bibinfo {pages}
  {017205} (\bibinfo {year} {2004})}\BibitemShut {NoStop}%
\bibitem [{\citenamefont {Jarillo-Herrero}\ \emph {et~al.}(2005)\citenamefont
  {Jarillo-Herrero}, \citenamefont {Kong}, \citenamefont {van~der Zant},
  \citenamefont {Dekker}, \citenamefont {Kouwenhoven},\ and\ \citenamefont
  {Franceschi}}]{Jarillo-Herrero:2005Nat}%
  \BibitemOpen
  \bibfield  {author} {\bibinfo {author} {\bibfnamefont {P.}~\bibnamefont
  {Jarillo-Herrero}}, \bibinfo {author} {\bibfnamefont {J.}~\bibnamefont
  {Kong}}, \bibinfo {author} {\bibfnamefont {H.~S.}\ \bibnamefont {van~der
  Zant}}, \bibinfo {author} {\bibfnamefont {C.}~\bibnamefont {Dekker}},
  \bibinfo {author} {\bibfnamefont {L.~P.}\ \bibnamefont {Kouwenhoven}}, \ and\
  \bibinfo {author} {\bibfnamefont {S.~D.}\ \bibnamefont {Franceschi}},\ }\href
  {\doibase 10.1038/nature03422} {\bibfield  {journal} {\bibinfo  {journal}
  {Nature}\ }\textbf {\bibinfo {volume} {434}},\ \bibinfo {pages} {484}
  (\bibinfo {year} {2005})}\BibitemShut {NoStop}%
\bibitem [{\citenamefont {Nishida}(2013)}]{Nishida:2013PRL}%
  \BibitemOpen
  \bibfield  {author} {\bibinfo {author} {\bibfnamefont {Y.}~\bibnamefont
  {Nishida}},\ }\href {\doibase 10.1103/PhysRevLett.111.135301} {\bibfield
  {journal} {\bibinfo  {journal} {Phys. Rev. Lett.}\ }\textbf {\bibinfo
  {volume} {111}},\ \bibinfo {pages} {135301} (\bibinfo {year} {2013})},\
  \Eprint {http://arxiv.org/abs/1308.3208} {arXiv:1308.3208
  [cond-mat.quant-gas]} \BibitemShut {NoStop}%
\bibitem [{\citenamefont {Nishida}(2016)}]{Nishida:2016PRA}%
  \BibitemOpen
  \bibfield  {author} {\bibinfo {author} {\bibfnamefont {Y.}~\bibnamefont
  {Nishida}},\ }\href {\doibase 10.1103/PhysRevA.93.011606} {\bibfield
  {journal} {\bibinfo  {journal} {Phys. Rev.}\ }\textbf {\bibinfo {volume}
  {A93}},\ \bibinfo {pages} {011606} (\bibinfo {year} {2016})},\ \Eprint
  {http://arxiv.org/abs/1508.07098} {arXiv:1508.07098 [cond-mat.quant-gas]}
  \BibitemShut {NoStop}%
\bibitem [{\citenamefont {Yasui}\ and\ \citenamefont
  {Sudoh}(2013)}]{Yasui:2013xr}%
  \BibitemOpen
  \bibfield  {author} {\bibinfo {author} {\bibfnamefont {S.}~\bibnamefont
  {Yasui}}\ and\ \bibinfo {author} {\bibfnamefont {K.}~\bibnamefont {Sudoh}},\
  }\href {\doibase 10.1103/PhysRevC.88.015201} {\bibfield  {journal} {\bibinfo
  {journal} {Phys. Rev.}\ }\textbf {\bibinfo {volume} {C88}},\ \bibinfo {pages}
  {015201} (\bibinfo {year} {2013})},\ \Eprint {http://arxiv.org/abs/1301.6830}
  {arXiv:1301.6830 [hep-ph]} \BibitemShut {NoStop}%
\bibitem [{\citenamefont {Hattori}\ \emph {et~al.}(2015)\citenamefont
  {Hattori}, \citenamefont {Itakura}, \citenamefont {Ozaki},\ and\
  \citenamefont {Yasui}}]{Hattori:2015hka}%
  \BibitemOpen
  \bibfield  {author} {\bibinfo {author} {\bibfnamefont {K.}~\bibnamefont
  {Hattori}}, \bibinfo {author} {\bibfnamefont {K.}~\bibnamefont {Itakura}},
  \bibinfo {author} {\bibfnamefont {S.}~\bibnamefont {Ozaki}}, \ and\ \bibinfo
  {author} {\bibfnamefont {S.}~\bibnamefont {Yasui}},\ }\href {\doibase
  10.1103/PhysRevD.92.065003} {\bibfield  {journal} {\bibinfo  {journal} {Phys.
  Rev.}\ }\textbf {\bibinfo {volume} {D92}},\ \bibinfo {pages} {065003}
  (\bibinfo {year} {2015})},\ \Eprint {http://arxiv.org/abs/1504.07619}
  {arXiv:1504.07619 [hep-ph]} \BibitemShut {NoStop}%
\bibitem [{\citenamefont {Ozaki}\ \emph {et~al.}(2015)\citenamefont {Ozaki},
  \citenamefont {Itakura},\ and\ \citenamefont {Kuramoto}}]{Ozaki:2015sya}%
  \BibitemOpen
  \bibfield  {author} {\bibinfo {author} {\bibfnamefont {S.}~\bibnamefont
  {Ozaki}}, \bibinfo {author} {\bibfnamefont {K.}~\bibnamefont {Itakura}}, \
  and\ \bibinfo {author} {\bibfnamefont {Y.}~\bibnamefont {Kuramoto}},\ }\href
  {\doibase 10.1103/PhysRevD.94.074013} {\bibfield  {journal} {\bibinfo
  {journal} {Phys. Rev.}\ }\textbf {\bibinfo {volume} {D94}},\ \bibinfo {pages}
  {074013} (\bibinfo {year} {2015})},\ \Eprint
  {http://arxiv.org/abs/1509.06966} {arXiv:1509.06966 [hep-ph]} \BibitemShut
  {NoStop}%
\bibitem [{\citenamefont {Yasui}(2016{\natexlab{a}})}]{Yasui:2016ngy}%
  \BibitemOpen
  \bibfield  {author} {\bibinfo {author} {\bibfnamefont {S.}~\bibnamefont
  {Yasui}},\ }\href {\doibase 10.1103/PhysRevC.93.065204} {\bibfield  {journal}
  {\bibinfo  {journal} {Phys. Rev.}\ }\textbf {\bibinfo {volume} {C93}},\
  \bibinfo {pages} {065204} (\bibinfo {year} {2016}{\natexlab{a}})},\ \Eprint
  {http://arxiv.org/abs/1602.00227} {arXiv:1602.00227 [hep-ph]} \BibitemShut
  {NoStop}%
\bibitem [{\citenamefont {Yasui}\ \emph {et~al.}(2016)\citenamefont {Yasui},
  \citenamefont {Suzuki},\ and\ \citenamefont {Itakura}}]{Yasui:2016svc}%
  \BibitemOpen
  \bibfield  {author} {\bibinfo {author} {\bibfnamefont {S.}~\bibnamefont
  {Yasui}}, \bibinfo {author} {\bibfnamefont {K.}~\bibnamefont {Suzuki}}, \
  and\ \bibinfo {author} {\bibfnamefont {K.}~\bibnamefont {Itakura}},\
  }\href@noop {} {\  (\bibinfo {year} {2016})},\ \Eprint
  {http://arxiv.org/abs/1604.07208} {arXiv:1604.07208 [hep-ph]} \BibitemShut
  {NoStop}%
\bibitem [{\citenamefont {Yasui}\ and\ \citenamefont
  {Sudoh}(2016)}]{Yasui:2016hlz}%
  \BibitemOpen
  \bibfield  {author} {\bibinfo {author} {\bibfnamefont {S.}~\bibnamefont
  {Yasui}}\ and\ \bibinfo {author} {\bibfnamefont {K.}~\bibnamefont {Sudoh}},\
  }\href {\doibase 10.1103/PhysRevC.95.035204} {\bibfield  {journal} {\bibinfo
  {journal} {Phys. Rev.}\ }\textbf {\bibinfo {volume} {C95}},\ \bibinfo {pages}
  {035204} (\bibinfo {year} {2016})},\ \Eprint
  {http://arxiv.org/abs/1607.07948} {arXiv:1607.07948 [hep-ph]} \BibitemShut
  {NoStop}%
\bibitem [{\citenamefont {Yasui}(2016{\natexlab{b}})}]{Yasui:2016yet}%
  \BibitemOpen
  \bibfield  {author} {\bibinfo {author} {\bibfnamefont {S.}~\bibnamefont
  {Yasui}},\ }\href@noop {} {\  (\bibinfo {year} {2016}{\natexlab{b}})},\
  \Eprint {http://arxiv.org/abs/1608.06450} {arXiv:1608.06450 [hep-ph]}
  \BibitemShut {NoStop}%
\bibitem [{\citenamefont {Kanazawa}\ and\ \citenamefont
  {Uchino}(2016)}]{Kanazawa:2016ihl}%
  \BibitemOpen
  \bibfield  {author} {\bibinfo {author} {\bibfnamefont {T.}~\bibnamefont
  {Kanazawa}}\ and\ \bibinfo {author} {\bibfnamefont {S.}~\bibnamefont
  {Uchino}},\ }\href {\doibase 10.1103/PhysRevD.94.114005} {\bibfield
  {journal} {\bibinfo  {journal} {Phys. Rev.}\ }\textbf {\bibinfo {volume}
  {D94}},\ \bibinfo {pages} {114005} (\bibinfo {year} {2016})},\ \Eprint
  {http://arxiv.org/abs/1609.00033} {arXiv:1609.00033 [cond-mat.str-el]}
  \BibitemShut {NoStop}%
\bibitem [{\citenamefont {Kimura}\ and\ \citenamefont
  {Ozaki}()}]{Kimura:2016inp}%
  \BibitemOpen
  \bibfield  {author} {\bibinfo {author} {\bibfnamefont {T.}~\bibnamefont
  {Kimura}}\ and\ \bibinfo {author} {\bibfnamefont {S.}~\bibnamefont {Ozaki}},\
  }\href@noop {} {\bibinfo  {journal} {in preparation}\ }\BibitemShut {NoStop}%
\bibitem [{\citenamefont {Macdonald}(1997)}]{Macdonald:1997}%
  \BibitemOpen
\bibfield  {journal} {  }\bibfield  {author} {\bibinfo {author} {\bibfnamefont
  {I.~G.}\ \bibnamefont {Macdonald}},\ }\href@noop {} {\emph {\bibinfo {title}
  {{Symmetric Functions and Hall Polynomials}}}},\ \bibinfo {edition} {2nd}\
  ed.\ (\bibinfo  {publisher} {Oxford University Press},\ \bibinfo {year}
  {1997})\BibitemShut {NoStop}%
\end{thebibliography}%

\end{document}